\documentclass[entropy,article,accept,pdftex,moreauthors]{Definitions/mdpi}
\usepackage{amsmath}
\usepackage{amssymb}
\usepackage{amsthm}
\usepackage{bbm}
\usepackage{color}
\usepackage{float}
\usepackage[T1]{fontenc}
\usepackage{graphicx}
\usepackage[utf8]{inputenc}
\usepackage{physics}
\usepackage{xcolor}
\newcommand{\revision}[1]{{\color{black}{#1}}}

\firstpage{1}
\pubvolume{27}
\issuenum{7}
\articlenumber{764}
\pubyear{2025}
\copyrightyear{2025}
\externaleditor{Leonardo dos Santos Lima}
\datereceived{28 May 2025}
\daterevised{5 July 2025}
\dateaccepted{16 July 2025}
\datepublished{18 July 2025}
\hreflink{https://doi.org/\linebreak  10.3390/e27070764}

\Title{Comparative Analysis of Robust Entanglement Generation in Engineered XX Spin Chains}

\TitleCitation{Comparative Analysis of Robust Entanglement Generation in Engineered XX Spin Chains}

\Author{{Eduardo K. Soares} $^{1,}$*\orcidA{}, {Gentil D.} de Moraes Neto $^{2,}${*}\orcidB{} and Fabiano M. Andrade $^{1,3,4,}${*}\orcidC{}}

\AuthorNames{Eduardo K. Soares, Gentil D. de Moraes Neto and Fabiano M. Andrade}
\AuthorCitation{{Soares, E.K.;} \linebreak  de Moraes Neto, G.D.; Andrade, F.M.}

\address{%
$^{1}$ \quad Programa de P\'os-Gradua\c{c}\~{a}o em Ci\^{e}ncias/F\'{i}sica, Universidade Estadual de Ponta Grossa,\linebreak  Ponta Grossa 84030-900, PR, Brazil \\
$^{2}$ \quad College of Physics and Engineering, Qufu Normal University, Qufu 273165, China\\
$^{3}$ \quad Departamento de Matem\'{a}tica e Estat\'{i}stica,
Universidade Estadual de Ponta Grossa,\linebreak  Ponta Grossa
84030-900, PR, Brazil\\
$^{4}$ \quad  Departamento de Física, Universidade Federal do Paraná, Curitiba 81531-980, PR, Brazil
}

\corres{Correspondence: edukso2002@gmail.com (E.K.S.); gdmneto@gmail.com (G.D.d.M.N.);\linebreak fmandrade@uepg.br (F.M.A.)}

\abstract{
We present a numerical investigation comparing two entanglement generation protocols in finite XX spin chains with varying spin magnitudes ($s = 1/2, 1, 3/2 $). Protocol 1 (P1) relies on staggered couplings to steer correlations toward the ends of the chain. At the same time, Protocol 2 (P2) adopts a dual-port architecture that uses optimized boundary fields to mediate virtual excitations between terminal spins. Our results show that P2 consistently outperforms P1 in all spin values, generating higher-fidelity entanglement in shorter timescales when evaluated under the same system parameters. Furthermore, P2 exhibits superior robustness under realistic imperfections, including diagonal and off-diagonal disorder, as well as dephasing noise. \revision{{To} further assess the resilience of both protocols in experimentally relevant settings, we employ the pseudomode formalism to characterize the impact of non-Markovian noise on the entanglement dynamics. Our analysis reveals that the dual-port mechanism (P2) remains effective even when memory effects are present, as it reduces the excitation of bulk modes that would otherwise enhance environment-induced backflow.} Together, the scalability, efficiency, and noise resilience of the dual-port approach position it as a promising framework for entanglement distribution in solid-state quantum information platforms.
}


\begin{document}

\section{Introduction}

The advancement of quantum technologies depends crucially on the ability
to generate and control quantum resources in a reliable and scalable way~\cite{QT,Res}.
Among these resources, quantum entanglement stands out as a fundamental
component, enabling powerful protocols to operate within the limits of
local operations and classical communication (LOCC)~\cite{Res}.
Entangled states are at the heart of key quantum information processing
tasks, such as teleportation~\cite{tele}, sensing \cite{sensing}, and
communication \cite{Bennett_2014,PhysRevLett.96.010401}.
In particular, maximally entangled states, such as Bell pairs (EPR
states), are essential building blocks for these applications~\cite{einstein1935}.

Although significant advances have been made in generating entanglement
with photonic systems \cite{kwiat1995,hucul2015}, scalable solid-state
quantum processors require architectures that can produce and distribute
entanglement on demand, integrated directly with quantum registers
\cite{braunstein1998}.
Spin chains have thus emerged as promising candidates for short-range
quantum communication and entanglement distribution
\cite{bose2007,niko2004,kay2010}, due to their tunability and
compatibility with solid-state platforms.
Physical implementations span electron spins in quantum dots
\cite{damico2007}, magnetic molecules \cite{tejada2001}, and endohedral
fullerenes \cite{twamley2003}, making spin chains attractive as modular
elements for scalable quantum devices.

There is an extensive body of work exploring spin chains for quantum
state \mbox{transfer \cite{Bose_2003,Christandl_2004,bose2007},} entanglement
routing \cite{PhysRevA.66.032110,RevModPhys.80.517,Alsulami2022}, and
quantum bus architectures
\cite{yung2005,estarellas2017robust,Venuti2006,Venuti2007,
Giampaolo2009,niko2004_b,nikoBook}.
However, much of this research focuses on idealized spin-${1}/{2}$
models and often overlooks practical challenges such as disorder and
decoherence.
Additionally, the performance of higher-spin chains and comparisons
between different entanglement generation schemes remain largely
unexplored
\revision{
\cite{agarwal2024creatingtwoquditmaximallyentangled,PhysRevB.108.144414,
Refael_2007}}.

We address these gaps by comparing two entanglement generation protocols
based on XY-type spin chains:
Protocol 1 (P1), where alternating weak and strong couplings guide
quantum correlations toward the chain edges
\cite{Wilkinson2018,estarellas2017robust,estarellas2016,
huo2008,Venuti2007},
and Protocol 2 (P2), which employs symmetric perturbative couplings at
both ends to enhance the transport speed and facilitate the buildup of
quantum correlations
\revision{
\cite{neto2012nonlocal,yao2011robust,paganelli2013,almeida2019entanglement}.}
Using the \revision{XX} spin model, we systematically investigate the
entanglement dynamics for spins $s = {1}/{2}$, $1$, and ${3}/{2}$ under
both ideal and noisy conditions.
Our analysis is primarily based on extensive numerical simulations.
We quantify entanglement using the negativity measure, explicitly
including the effects of static disorder, both diagonal and
off-diagonal, as well as local decoherence channels within the model.
\revision{
Non-Markovian effects, originating from structured environments with
finite memory times or strong system--bath correlations, can significantly
influence quantum coherence, entanglement dynamics, and the fidelity of
quantum operations \cite{Rivas2014,shrikant2023quantum}.
Accurately capturing these effects is essential in understanding
realistic quantum devices.

To this end, we employ the pseudomode formalism
\cite{imamog1994stochastic,Garraway1997}, a conceptually transparent and
computationally efficient approach for the modeling of non-Markovian dynamics
induced by reservoirs with Lorentzian or near-Lorentzian spectral
densities.
In this framework, the environment is effectively replaced by a set of
auxiliary damped harmonic oscillators (pseudomodes) that interact
directly with the system.
This mapping enables the simulation of non-Markovian effects using
standard Lindblad master equations, thereby retaining essential memory
effects while remaining compatible with widely used \mbox{numerical
techniques}.}

Although effective Hamiltonians are derived to elucidate the underlying
physical mechanisms, our main conclusions are based entirely on the
numerical data.
Importantly, we observe that the dual-port databus protocol (P2) enables
faster and more robust entanglement generation compared to the staggered
(P1) scheme, especially in the presence of environmental noise and
fabrication imperfections.
To demonstrate practical applicability, we benchmark the entangling
times and parameter regimes against those accessible in the current
solid-state platforms.
Our results are relevant to a variety of systems—including trapped ions
\cite{Blatt2012}, superconducting qubit arrays \cite{Krantz2019},
nitrogen-vacancy centers in diamond~\cite{Doherty2013}, and quantum dot
devices \cite{Hanson2007}—where engineered spin--spin interactions and
coherent control have been experimentally realized.
This work thus paves the way for deterministic and scalable entanglement
sources in next-generation quantum processors and networks
\revision{
\cite{PhysRevA.81.032312,galve2010entanglement,PhysRevResearch.7.013268,
bayat2022entanglement}.}

\section{{Model} and Methods}

\begin{figure}[H]
  \includegraphics[width=0.7\columnwidth]{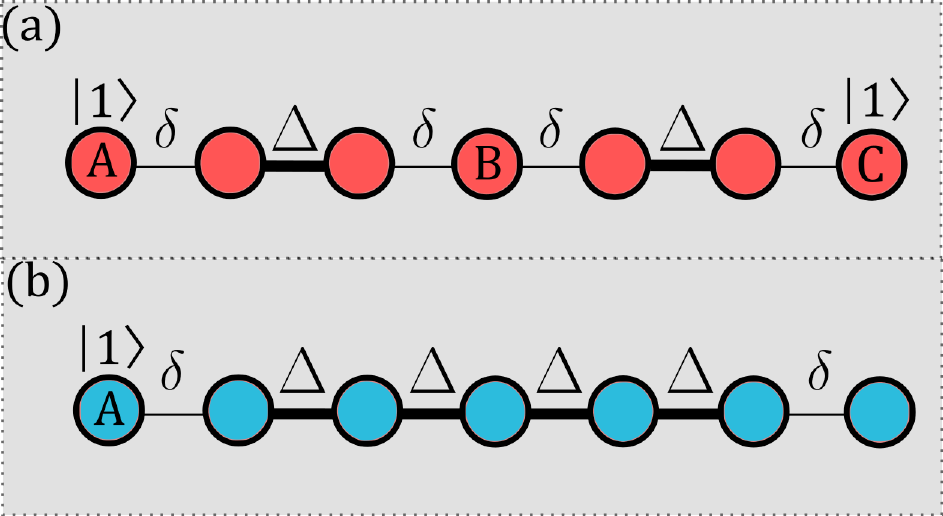}
  \caption{{(\textbf{a})} P1 and (\textbf{b}) P2 architectures.
    Bold lines represent $\Delta$ couplings, while thin lines indicate
    $\delta$ couplings.
  }
  \label{fig:fig1}
\end{figure}

We model the system as an $N$-site XX spin chain with {Hamiltonian}
\begin{equation}
H =
\sum_{i=1}^{N-1} J_i \left( S_i^x S_{i+1}^x + S_i^y S_{i+1}^y \right)
+ \sum_{i=1}^N B_i S_i^z,
\label{Ham}
\end{equation}
where $S_i^\alpha$ denotes spin-$s$ operators and $B_i$ denotes local magnetic
fields.
The coupling constant $J_i$ alternates between two values, $\Delta$ and
$\delta$, depending on the position in the chain (see
\mbox{Figure \ref{fig:fig1}).}

The system evolves under the Lindblad master equation
\cite{Lindblad1976,Gorini1976,Manzano2020}, which describes the
combined unitary and dissipative dynamics:
\begin{equation}
  \dot{\rho} = -i[H,\rho] +
  \gamma \sum_{i=1}^N
  \left( S_i^z \rho S_i^z - \frac{1}{2} { S_i^z S_i^z, \rho }
  \right).
  \label{ME}
\end{equation}
The dissipative term proportional to $\gamma$ introduces local pure
dephasing, a common and critical source of decoherence in quantum
systems.
This type of noise models the loss of quantum coherence without energy
exchange with the environment and is particularly relevant in platforms
such as superconducting qubits
\cite{Martinis2003,Schreier2008,Clarke2008,Krantz2019}, trapped ions
\cite{Schindler2013}, and ultracold atom simulators of spin chains
\cite{Bloch2008,Lewenstein2007}.
In such systems, fluctuations in the local environment or control
parameters often lead to dephasing noise that dominates other
dissipative processes.
Although we set $\gamma = 0$ unless otherwise stated to focus on the
coherent dynamics, we also consider finite $\gamma$ to assess the
robustness of quantum correlations under realistic conditions.

We assess entanglement between the chain ends via the negativity \cite{Peres1996,Horodecki1996,Vidal2002}, defined as
\begin{equation}
  \mathcal{N}(\hat{\rho})=
  \frac{\Vert\hat{\rho}^{T_{A}}\Vert_{1}-1}{2},
  \label{eq:neg}
\end{equation}
where $\hat{\rho}^{T_{A}}$ is the partial transpose of the quantum state
$\hat{\rho}$ with respect to subsystem A, and
$\Vert\hat{Y}\Vert_{1}=\tr|\hat{Y}|=\tr\sqrt{\hat{Y}^{\dagger}\hat{Y}}$
denotes the trace norm or the sum of the singular values of the
operator $\hat{Y}$. Alternatively, negativity can be calculated as
$\mathcal{N}(\hat{\rho})=\sum_i(|\varepsilon_i|-\varepsilon_i)/2$,
where $\varepsilon_i$ denotes the eigenvalues of the partially transposed
density matrix $\hat{\rho}$.
The maximum attainable value of $\mathcal{N}(\hat{\rho})$ is constrained
by the dimensionality of the Hilbert space, which depends on the spin
magnitude $s$.
Because we work across different dimensionality systems, we normalize
our negativity calculations relative to the theoretical maximum for the
specific spin value $s$, corresponding to the negativity of a maximally
entangled state in the relevant Hilbert space.

To assess whether the protocols generate a Bell state when the maximal
possible negativity $\mathcal{N}$ is achieved, we compute the fidelity
\cite{nielsen2010quantum}
\begin{align}
  F(\hat{\rho}, \hat{\sigma}) =
  \left(
  \tr \sqrt{ \sqrt{\hat{\rho}} \, \hat{\sigma} \, \sqrt{\hat{\rho}} }
  \right)
\end{align}
where $\hat{\rho}$ is the density matrix of the generated state, and
$\hat{\sigma}$ is the density matrix associated with the target state.
Fidelity of $F = 1$ indicates the perfect preparation of the target state,
while values below unity quantify deviations from this.

\subsection*{{Entanglement} Generation Protocols}
We investigate two different entanglement generation protocols,
illustrated in \mbox{Figure \ref{fig:fig1},} both designed to mediate
long-range entanglement between boundary spins in a finite chain.
Unlike traditional quantum communication setups focused on state
transfer fidelity, our objective here is the efficient and robust
creation of quantum correlations---specifically entanglement---between
distant parties.

P1 employs a staggered spin chain initialized in the state
\begin{equation}
    \ket{\psi(0)}=\ket{1}_A\otimes\ket{0}^{\otimes N-2}\otimes\ket{1}_C.
\end{equation}
Moreover, it evolves unitarily under the system Hamiltonian with
$B_i = 0$.
In this configuration, the boundary spins $A$ and $C$ are initially
excited, while the intermediate sites \revision{remain unexcited}.
The entanglement dynamics, in this case, arise from coherent exchange
interactions distributed across the entire chain.

For P2, the chain is initialized in the state
\begin{align}
    \ket{\psi(0)}=\ket{1}_A\otimes\ket{0}^{\otimes N-1},
\end{align}
but, here, a single excitation is localized at the sender \revision{(s)}
site, while all other spins, including the receiver \revision{(r)} at
the opposite end, begin in the unexcited state.
\revision{
For a spin-$s$ system, we define the computational basis states in terms
of the eigenstates of the $S_z$ operator.
Specifically, for spin-$1$, the basis states are
$\ket{0} \equiv \ket{m=-1}$, $\ket{1} \equiv \ket{m=0}$, and
$\ket{2} \equiv \ket{m=+1}$.
For spin-$3/2$, the definitions are $\ket{0} \equiv \ket{m=-3/2}$,
$\ket{1} \equiv \ket{m=-1/2}$, $\ket{2} \equiv \ket{m=+1/2}$, and
$\ket{3} \equiv \ket{m=+3/2}$.

Initial state preparation follows a consistent scheme across the
different spin systems.
In P1, the bulk spins are initialized in the minimal $S_z$ eigenstate,
$\ket{m=-s}$, corresponding to the lowest indexed basis state $\ket{0}$,
while the boundary spins are set to the maximal eigenstate,
$\ket{m=+s}$, corresponding to the highest indexed basis state.
In contrast, P2 initializes all spins uniformly in the
minimal eigenstate $\ket{m=-s}$.}
A zero magnetic field $B_i=0$ is applied in the bulk, while carefully
engineered, \textit{optimized boundary magnetic fields} are applied at
the extremities to enhance the coherent buildup of long-range
entanglement.

A central advantage of P2 is that the bulk (spins $2$ through $N-1$)
remains largely \revision{unexcited} during evolution.
That is, the intermediate spins undergo only \textit{{virtual}
  excitation}, which avoids a significant population of the bulk and
enables the boundary spins to interact effectively as if they were
directly coupled.
This virtual coupling mechanism reduces the influence of imperfections
within the chain, such as diagonal and off-diagonal disorder or local
dephasing, thereby supporting the robust generation of entanglement
between the sender and receiver.

Although structurally reminiscent of state transfer protocols, the goal
here is not to maximize the transfer fidelity but to exploit coherent
dynamics for the fast and resilient generation of entanglement.
This distinction is central to our investigation, and we provide
detailed numerical results in the following section to validate the
effectiveness of both protocols under various conditions.
In particular, we systematically compare their performance across
different spin magnitudes, analyze their resilience to disorder and
decoherence, and explore how the introduction of site-dependent magnetic
fields ($B_i \neq 0$) affects entanglement generation.

The analysis reveals that P2 offers three key advantages over P1:
(i) it achieves maximal entanglement between boundary spins on shorter
timescales, with this effect being particularly pronounced in the
spin-${1}/{2}$ case;
(ii) it demonstrates enhanced robustness against both diagonal and
off-diagonal disorder, as well as dephasing noise;
and
(iii) it maintains high entanglement generation efficiency even in
higher-spin systems ($s = 1$ and $3/2$), where P1 shows reduced
effectiveness.

These benefits stem from the engineered boundary control and the
architecture's ability to harness virtual excitations for indirect but
coherent boundary coupling, effectively bypassing the detrimental
effects of bulk-mediated decoherence.
The validity of these claims and the quantitative characterization of
these mechanisms will be fully explored in the next section.
\revision{
We note that various mechanisms have been explored in the literature for
the generation of long-distance entanglement in spin chains, providing
complementary approaches to the coherent transfer-based protocols that we
analyze.
Static methods, such as engineered couplings in dimerized chains and
entanglement routers \cite{Venuti2006,campos2007quantum}, can create
ground states with naturally entangled boundaries.
Dissipative approaches
\cite{verstraete2009quantum,lin2013dissipative,dias2023entanglement}
utilize controlled environmental coupling to prepare entangled steady
states.
More recent developments include measurement-based post-selection
\cite{wallraff2020heralded} and the exploitation of topologically
protected edge modes \cite{bahri2015localization}.
Additionally, numerous protocols for high-fidelity quantum information
transfer \cite{nikoBook} could potentially be adapted to
maximize boundary entanglement.
While we do not provide a quantitative comparison here, these diverse
methods highlight the rich landscape of available techniques and
position our protocols within this broader research context.
}

\section{Results}

\subsection{Benchmark Without Noise: Dynamics in Pristine Chains}

\begin{figure}[H]
  \includegraphics[width=.7\columnwidth]{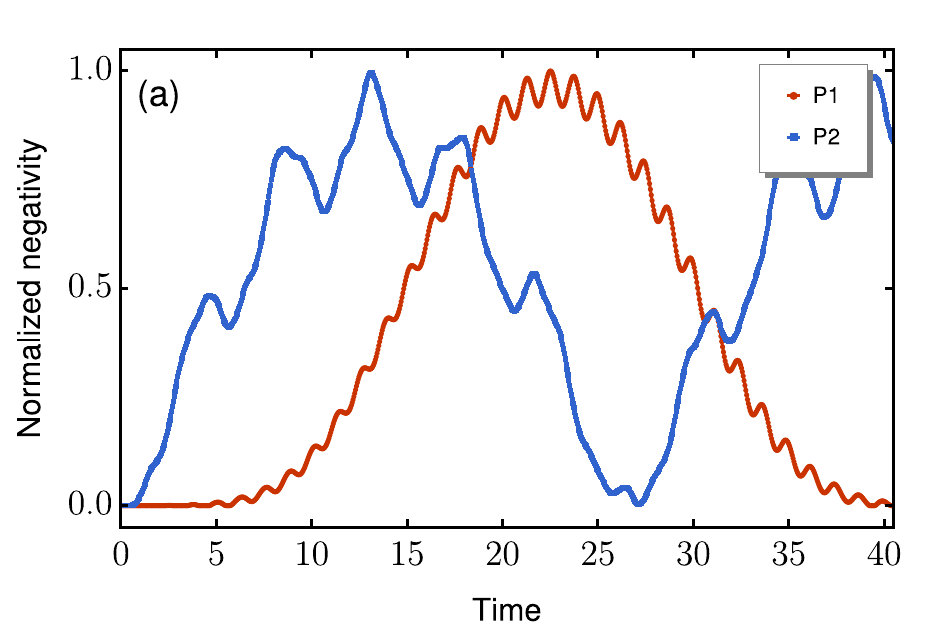}\\
  \includegraphics[width=.7\columnwidth]{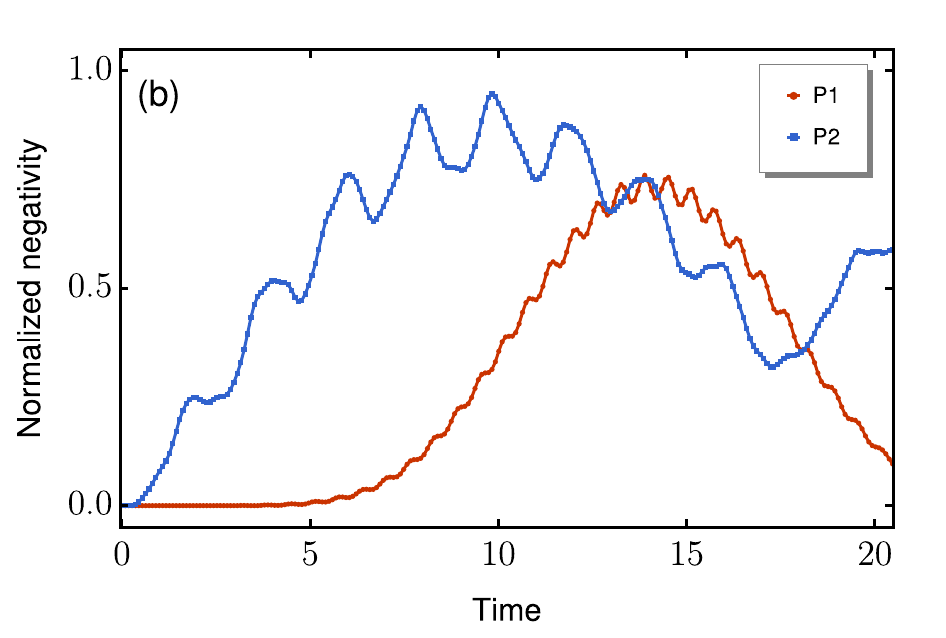}\\
  \includegraphics[width=.7\columnwidth]{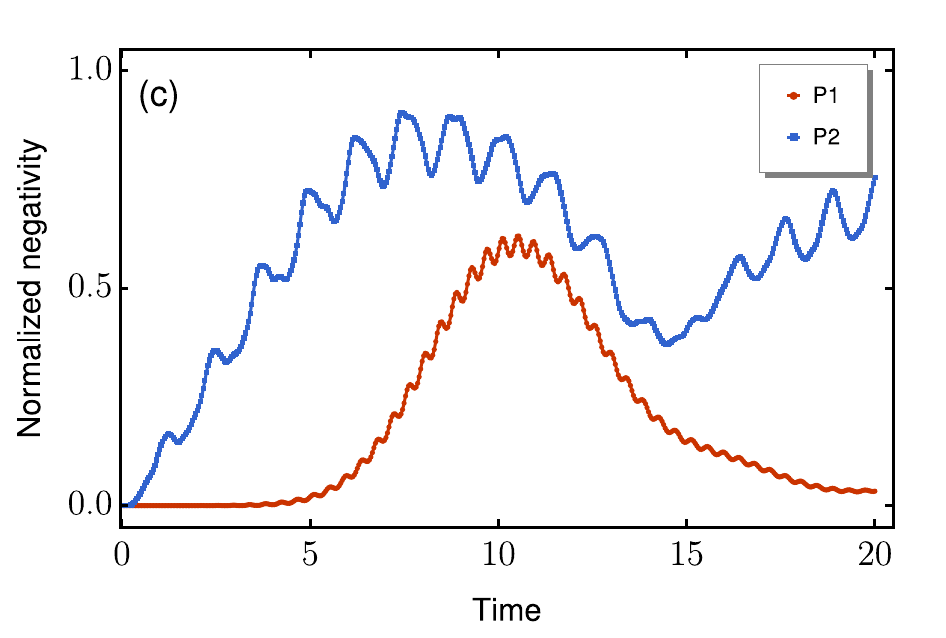}
  \caption{
    Time evolution of the end‑to‑end negativity for (\textbf{a}) $s=1/2$,
    (\textbf{b}) $s=1$, and (\textbf{c}) $s=3/2$. All traces correspond to the same
    dimerization ratio $\Delta/\delta = 10$.
    Results are shown for P1 (red curves) and P2 (blue curves).
  }
  \label{fig:fig2}
\end{figure}

To establish a baseline, we first compare the coherent dynamics of the
two architectures in the absence of disorder or decoherence for $N=7$.
Figure \ref{fig:fig2} shows the time evolution of the end-to-end
negativity for spin magnitudes $s={1}/{2}$, $1$, and ${3}/{2}$.
The results demonstrate that P2 reaches its first entanglement maximum
significantly more quickly than P1 while maintaining robust performance across
different spin dimensions.
The quantitative data extracted from these curves are presented in Table
\ref{tab:table1}.

\begin{table}[H]
  \caption{{Peak} end‑to‑end negativity $\mathcal N$ and corresponding evolution
    time $t$ for P1 (subscript $1$)  and P2 (subscript $2$).
    Times are in units of the weak coupling $\delta$;
    $B_A=B_B=B$ are the optimal boundary~fields.}
    \begin{tabularx}{\textwidth}{CCCCCC}
    \toprule
      \textbf{Spin}  & \boldmath{$\mathcal {N}_{1}$} & \boldmath{$t_{1}$} & \boldmath{$\mathcal{N}_{2}$} & \boldmath{$t_{2}$}
      & \boldmath{$B$} \\
      \midrule
      $1/2$ & 1    & 22.50$\delta$ & 1    & 13$\delta$   & 3.7 \\
      $1$   & 0.75 & 13.9$\delta$  & 0.94 & 9.8$\delta$  & 2.9 \\
      $3/2$ & 0.62 & 10.54$\delta$ & 0.9  & 7.45$\delta$ & 4.7\\
      \bottomrule
    \end{tabularx}
    \label{tab:table1}
\end{table}

\begin{figure}[H]
  \includegraphics[width=.7\columnwidth]{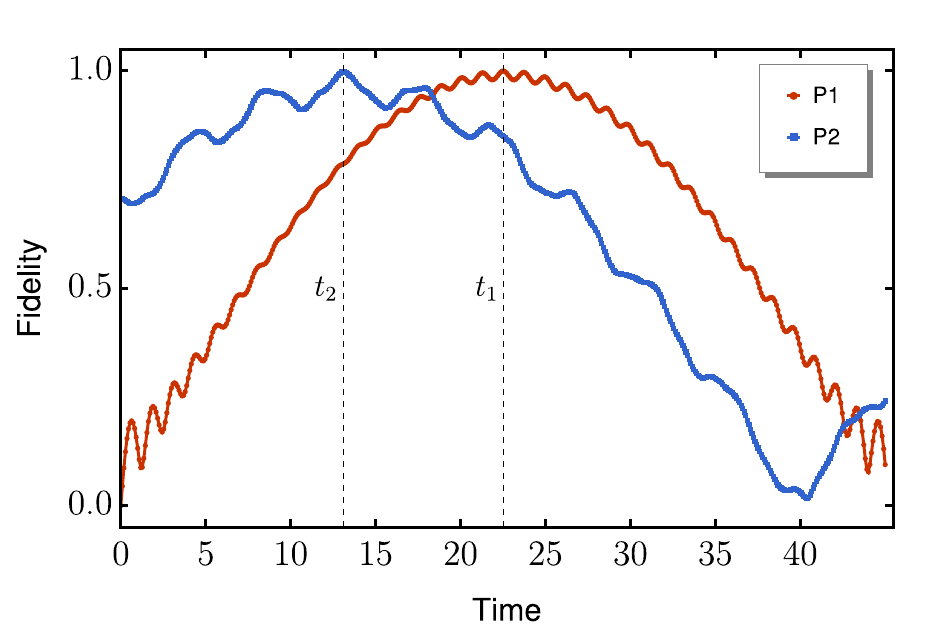}
  \caption{
    Time evolution of fidelity when considering $\ket{\psi^+}$ as the
    target state. The red line represents P1, while the blue line
    represents P2.
    The time at which maximal entanglement is achieved for each protocol
    is marked with a dashed vertical line.
    \revision{We set the dimerization ratio to $\Delta/\delta = 10$.}
  }
  \label{fig:fig3}
\end{figure}

In all cases, P2 exhibits faster entanglement generation and
consistently achieves higher negativity values.
In particular, for $s=1/2$, both protocols output the Bell state
$\ket{\psi^+} = (\ket{01} + \ket{10})/\sqrt{2}$ when maximally
entangled, although P2 requires a local $-\pi/2$ rotation to be applied on
the $z$-axis of the qubit at site $N$, via the application of the $R_z$
gate.
The fidelity dynamics concerning the target state $\ket{\psi^+}$ are
shown in Figure \ref{fig:fig3}, where the necessary rotation for
P2 has already been applied before the fidelity calculation.
\vspace{-9pt}

\begin{figure}[H]
  \includegraphics[width=0.7\columnwidth]{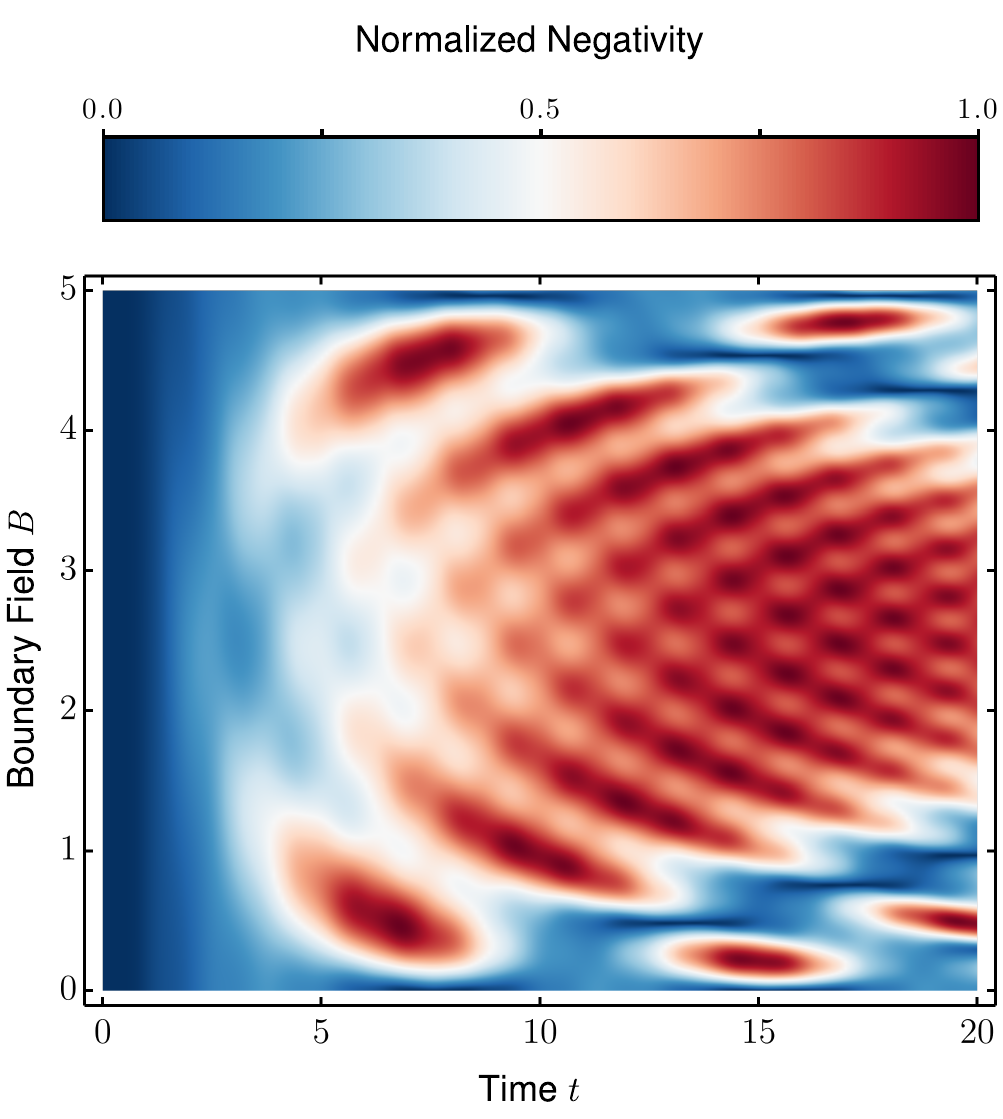}
  \caption{
    Contour plot of negativity values as a function of time and the
    magnetic field applied to the boundaries for a $N=7$ chain.
    \revision{We set the dimerization ratio to $\Delta/\delta = 10$.}
  }
  \label{fig:fig4}
\end{figure}

P2 can also be extended to arbitrary chain lengths $N$ when
$s=1/2$, making it possible to obtain maximally entangled states for
higher values of $N$.
For each system size, we only need to optimize the boundary magnetic
field $B$ to find the maximal possible negativity between terminal
spins.
This simple adjustment of $B$ for different $N$ consistently yields
maximal or near-maximal entanglement, demonstrating the scalability of
the protocol.
To determine the optimal value of $B$, we numerically analyze the
relationship between $B$ and the resulting negativity, identifying the
parameter regimes and time scales that maximize the negativity, as shown in
Figure \ref{fig:fig4}.

\vspace{-12pt}

\vspace{-12pt}

\subsection{Robustness of the Spin‑{1/2} Protocol}

The performance advantage of P2 is most relevant when it survives
realistic imperfections.
Therefore, we investigate its stability against static disorder,
focusing on the spin‑$1/2$ chain as a representative and experimentally
accessible platform.
To achieve this, we quantitatively assess its stability by introducing
static disorder in spin-${1}/{2}$ chains—a platform chosen for both its
theoretical tractability and its experimental relevance.
The analysis focuses on two fundamental channels of disorder that
reflect distinct physical~origins.

First, \textit{{on-site} (diagonal) disorder} tests P2's sensitivity to
variations in the fine-tuned boundary magnetic fields essential for its
operation.
To model fabrication-induced energy offsets, we introduce random local
fields $h_i = E d_i\delta$ with $d_i \in [-0.5, 0.5]$ uniformly
distributed, where $E$ scales the disorder strength relative to weak
coupling $\delta$.
This modifies the Hamiltonian as
\begin{equation}
  H \rightarrow H + E\delta(d_1 S_1^z + d_{N-1} S_{N-1}^z).
\end{equation}
For comparison, we apply identical perturbations to P1, establishing a
performance baseline under equivalent conditions.

Second, \textit{{coupling} (off-diagonal) disorder} captures imperfections
in exchange interactions arising from material defects or control
errors. The modified couplings $J_i \rightarrow J_i + E d_i \delta$
yield the adjusted Hamiltonian
\begin{equation}
  H \rightarrow
  \sum_{i=1}^{N-1} (J_i + E d_i \delta)
  \left(S_i^x S_{i+1}^x + S_i^yS_{i+1}^y\right).
\end{equation}

We investigate three distinct regimes:
(i) pure on-site disorder, (ii) pure coupling disorder, and (iii)
simultaneous disorder.
For each disorder strength $E$, we ensure statistical reliability by
computing the end-to-end peak negativity across $10^3$ realizations.
Specifically, for each realization, we simulate the time evolution under
the corresponding disordered Hamiltonian, record the maximum
entanglement value attained during the evolution, and then average this
value over all realizations.
The resulting data points, plotted for varying $E$, quantify the
robustness of both protocols against realistic experimental
imperfections.

\begin{figure}[H]
  \includegraphics[width=.7\columnwidth]{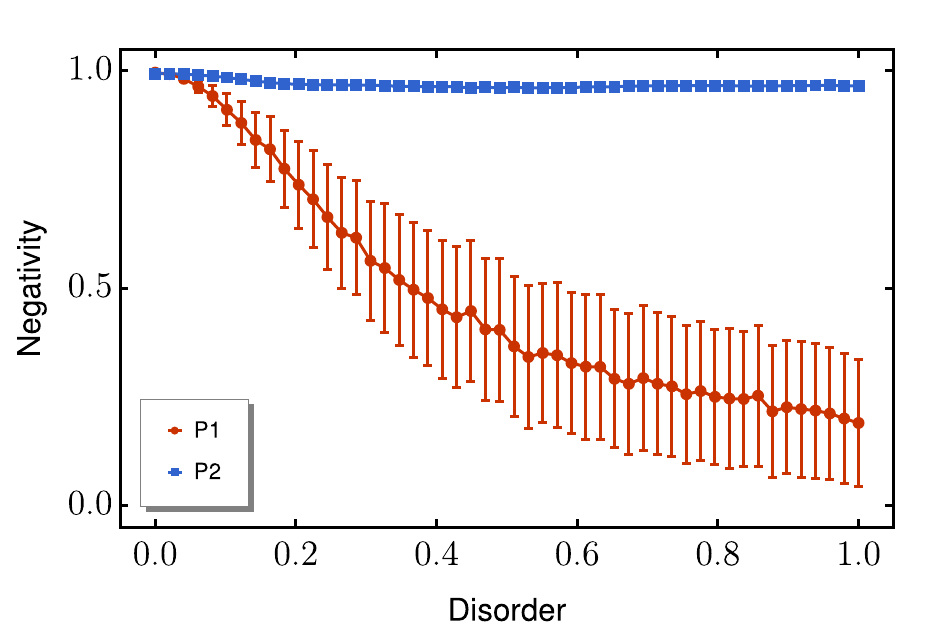}
  \caption{
    Average peak negativity as function of diagonal disorder strength
    $E$.
    The red line represents P1, while P2 is shown as a blue line.
    The red and blue bars indicate the standard deviation from the mean
    for each protocol.
    \revision{We set the dimerization ratio to $\Delta/\delta = 10$.}
  }
  \label{fig:fig5}
\end{figure}

Three cases are analyzed:
(i) pure on‑site disorder,
(ii) pure coupling disorder, and
(iii) both disorder types acting simultaneously.
We begin by analyzing pure on-site disorder.
Figure \ref{fig:fig5} shows that P2 maintains excellent
performance even at high disorder strengths, while P1 suffers
significant entanglement degradation.
This robustness is particularly valuable for practical implementations,
as it allows for high entanglement generation (high negativity values)
despite imperfections in the applied boundary magnetic fields.

\begin{figure}[H]
  \includegraphics[width=.7\columnwidth]{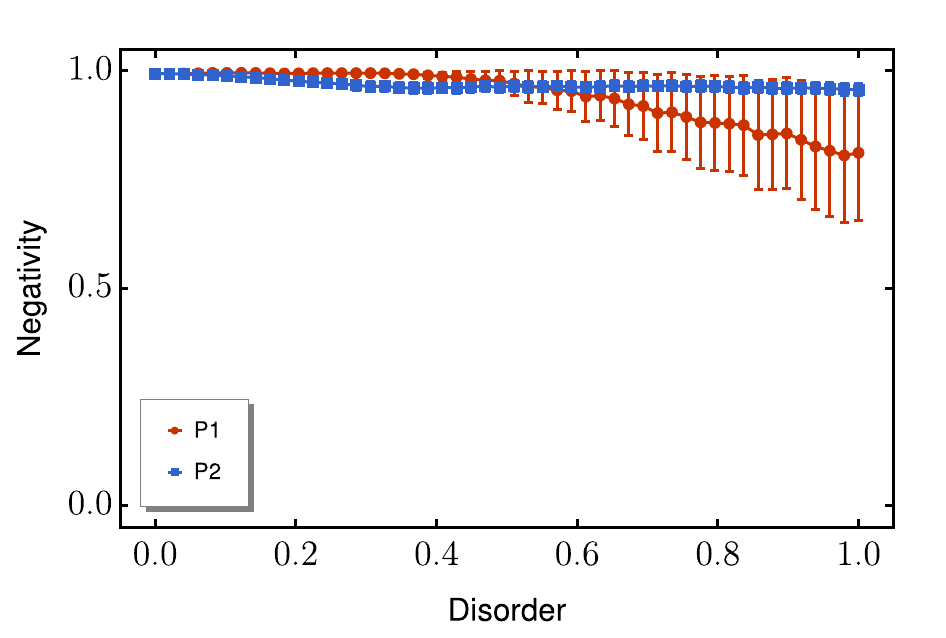}
  \caption{
    Average peak negativity as function of off-diagonal disorder
    strength $E$.
    The red line represents P1, while P2 is shown as a blue line.
    The red and blue bars indicate the standard deviation from the mean
    for each protocol.
    \revision{We set the dimerization ratio to $\Delta/\delta = 10$.}
  }
  \label{fig:fig6}
\end{figure}

\begin{figure}[H]
  \includegraphics[width=.7\columnwidth]{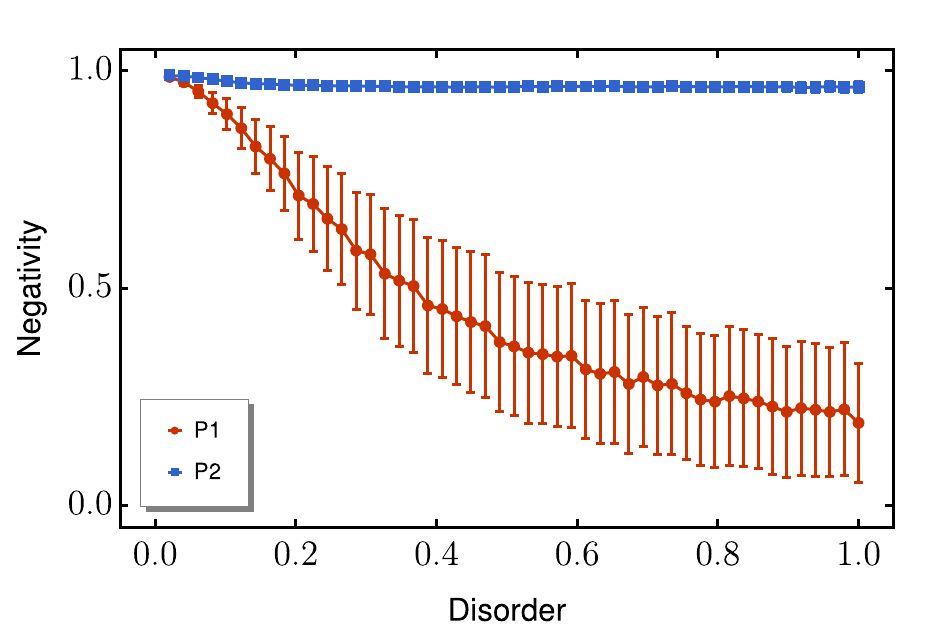}
  \caption{{Average} peak negativity as function of combined diagonal and
    off-diagonal disorder strength $E$.
    The red line represents P1, while P2 is shown as a blue line.
    The red and blue bars indicate the standard deviation from the mean
    for each protocol.
    \revision{ We set the dimerization ratio to $\Delta/\delta = 10$.}
  }
  \label{fig:fig7}
\end{figure}

A similar advantage emerges for coupling disorder
(Figure \ref{fig:fig6}), where P2 maintains substantial
entanglement ($\mathcal{N} \approx 0.8$) even at values of
$E\approx 0.75\delta$.
Interestingly, when diagonal and off-diagonal disorder are present
simultaneously, P2 continues to outperform P1, as shown in
Figure \ref{fig:fig7}.
This consistent superiority across all disorder regimes confirms P2's
exceptional resilience to typical solid-state fabrication imperfections.


\vspace{-6pt}

\subsection{Resistance to Dephasing}

\begin{figure}[H]
  \includegraphics[width=0.7\columnwidth]{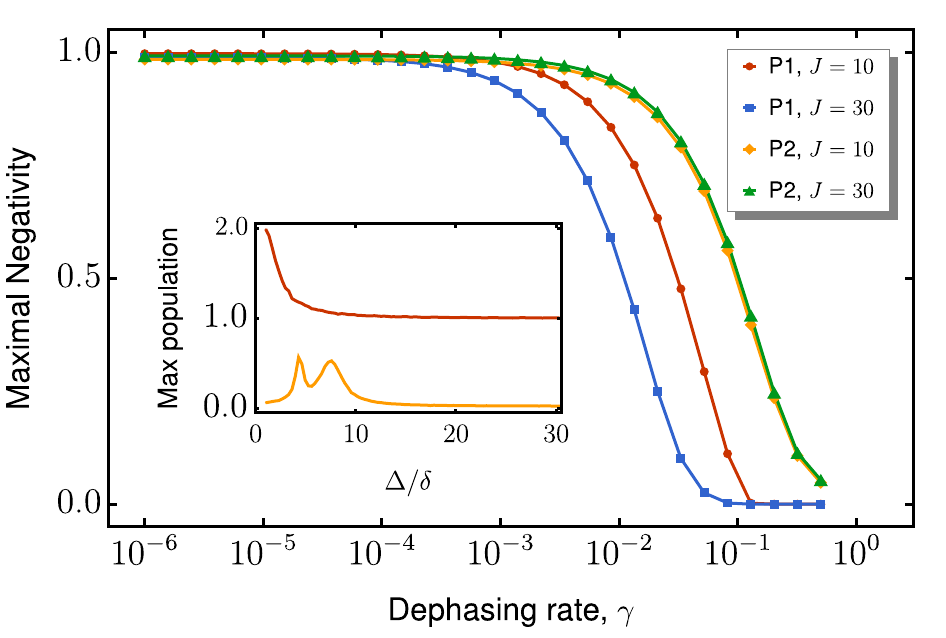}
  \caption{
    Peak end-to-end negativity as function of boundary dephasing
    rate $\gamma$, shown for two coupling regimes: $\Delta = 10$ and
    $\Delta = 30$, with $\delta = 1$.
    The main plot compares the performance of protocols P1 and P2,
    highlighting the enhanced robustness of P2, which exhibits a slower
    decay in entanglement under increasing dephasing.
    The inset displays the maximum population in the bulk channel for
    each protocol, demonstrating that P2 maintains significantly lower
    excitation in the intermediate spins across both coupling regimes.
  }
  \label{fig:fig8}
\end{figure}

A ubiquitous source of decoherence in solid-state devices is the pure
dephasing of the qubits that terminate the spin chain and interface it
with external control circuitry.
To assess its influence, we numerically evolve the \emph{full} Lindblad
master  equation [see \mbox{Equation \eqref{ME}}], in which the term
\begin{equation}
\gamma\sum_{i=1}^{N}\!\Bigl(S_i^{z}\hat{\rho} S_i^{z} - \frac{1}{2}\{(S_i^{z})^{2},\hat{\rho}\}\Bigr),
\end{equation}
already models local dephasing for every site.
By sweeping the dephasing rate $\gamma$ and recording the peak
end-to-end negativity, we obtain the curves in Figure \ref{fig:fig8}.
The qualitative difference in the decay trends between protocols P1 and P2
can be understood through the lens of the effective Hamiltonian derived
in Appendix \ref{app:effective_dynamics}.


\vspace{-12pt}

\revision{
In the dispersive regime ($\delta \ll \Delta$), and following
well-established perturbative techniques
\cite{Klimov2002master,Lee2018}, the spin chain dynamics underlying P2
can be effectively reduced to a two-qubit model,}
\begin{equation}
  H_{\mathrm{eff}} =
  \chi\bigl(S^{+}_{e}S^{-}_{r} + S^{-}_{e}S^{+}_{r}\bigr), \qquad
  \chi = \sum_{k}\frac{\bar\lambda_{k}^{2}}{\zeta_{k}},
\end{equation}
with an accompanying Lindblad term
\begin{equation}
\dot\rho_{\mathrm{eff}} = \frac{\Gamma}{2}\!\sum_{j=e,r}\! \bigl(2S^{z}_{j}\rho_{\mathrm{eff}}S^{z}_{j} - \{(S^{z}_{j})^{2},\rho_{\mathrm{eff}}\}\bigr),
\end{equation}
and $\lambda_k$ defined as in Equation \eqref{eq:lambdak}.
Here, the dephasing rate $\Gamma$ is renormalized as
\begin{equation}
\Gamma = \gamma\sum_{k}\frac{\bar\lambda_{k}^{2}}{\zeta_{k}^{2}}
\propto \gamma\left(\frac{\lambda}{J}\right)^{2},
\end{equation}
indicating that dephasing within the bulk enters only at higher order.
Since the chain spins remain virtually unexcited, the entangled state
decays predominantly through the suppressed rate $\Gamma$, explaining
the gradual decline in negativity in P2 (Figure \ref{fig:fig8}).
In contrast, P1 lacks this protection, leading to more pronounced
sensitivity to dephasing.

This distinction is clearly illustrated in the inset of
Figure \ref{fig:fig8}, which plots the maximum bulk excitation as a
function of the coupling ratio $\Delta/\delta$ for both protocols.
P2’s enhanced resilience to dephasing arises from its effective
decoupling from the bulk, whereas P1 relies on direct excitation
transport through the chain, making it significantly more vulnerable to
noise.
The key difference lies in how entanglement is generated: P1 requires
the physical propagation of excitations through intermediate spins
before boundary entanglement can be established.
As each of these intermediate spins becomes populated, the system
accumulates dephasing noise at each site.
This sequential exposure results in a larger bulk population (see the
red curve in the inset of Figure \ref{fig:fig8}) and sharper
degradation in entanglement.
In contrast, P2 consistently maintains low bulk occupation (orange
curve in the inset of Figure \ref{fig:fig8}) across all tested
\(\Delta/\delta\) ratios.
This population suppression directly explains the significantly flatter
negativity decay curve for P2 in the main panel of
\mbox{Figure \ref{fig:fig8}}: by avoiding the buildup of noise along the
chain, it preserves the entanglement more effectively under dephasing.

The enhanced protection in P2 comes from two key factors related to the
dimerization ratio $\Delta/\delta$.
First, higher values push the system deeper into the dispersive regime,
suppressing the real chain occupation.
Second, the effective dephasing rate $\Gamma \propto
\gamma(\lambda/J)^2$ decreases quadratically with $\zeta_{k}$, explaining
why P2's negativity curves in Figure \ref{fig:fig8} decay
increasingly slowly as $\Delta/\delta$ grows.
The combination of these effects, i.e., the minimal bulk population and
suppressed $\Gamma$, gives P2 its characteristically flat negativity
decay.

In contrast, P1's excitation-mediated transport remains fundamentally
exposed to dephasing regardless of $\Delta/\delta$, as its physical
propagation mechanism inevitably populates intermediate sites.
While stronger dimerization may slightly reduce bulk occupation, it
cannot eliminate the accumulation of sequential noise along the chain.
This stark difference highlights the central advantage of virtual
tunneling: by avoiding real excitations in the bulk, P2 naturally
decouples from noise sources while maintaining efficient end-to-end
entanglement generation.

\revision{
\subsection{Non-Markovian Effects}

The accurate description of open quantum systems is essential for the
development and operation of noisy intermediate-scale quantum (NISQ)
devices \cite{Preskill2018}.
In these devices, the interaction with the environment often leads to
decoherence and dissipation that cannot be fully captured by memoryless
(Markovian) approximations.
Non-Markovian effects, which arise due to structured environments with
significant memory times or strong system--bath correlations, can
substantially influence quantum coherence, the entanglement dynamics, and
ultimately the fidelity of quantum operations
\cite{Rivas2014,shrikant2023quantum}.
For example, in superconducting qubits, non-Markovian noise from
two-level system defects leads to coherence revivals \cite{Muller2015},
while, in quantum dots, phonon-induced memory effects cause
a non-exponential decay in entanglement \cite{Cywiński2009}.

Several theoretical and numerical approaches have been developed to
model non-Markovian dynamics.
These include hierarchical equations of motion (HEOM)
\cite{Tanimura2006}, time-convolutionless and Nakajima--Zwanzig master
equations \cite{Breuer2002}, stochastic Schrödinger equations with
colored noise \cite{Diosi1998}, and chain mapping techniques
\cite{Chin2010}.
Each of these methods presents trade-offs between accuracy,
computational complexity, and ease of implementation, with HEOM being
numerically exact but computationally expensive for large systems, while
stochastic methods offer flexibility but require ensemble averaging.

Among these, the \emph{{pseudomode} formalism} offers a conceptually
clear and numerically efficient framework for the simulation of non-Markovian
effects arising from reservoirs characterized by Lorentzian or
near-Lorentzian spectral densities
\cite{imamog1994stochastic,Garraway1997}.
The formalism effectively replaces the structured environment with one or
more auxiliary damped harmonic oscillators (pseudomodes) coupled
directly to the system. The resulting dynamics can then be simulated
using standard Lindblad master equations, thereby leveraging
well-established numerical solvers while capturing essential memory
effects.


Consider a general open quantum system described by the system
Hamiltonian $H_S$ interacting with a bosonic reservoir.
The total Hamiltonian of the system plus bath is
\begin{equation}
  H_{\mathrm{tot}} = H_S + \sum_k \omega_k b_k^\dagger b_k +
  \sum_k \left( g_k L b_k^\dagger + g_k^* L^\dagger b_k \right),
\end{equation}
where $L$ is a system operator coupling to the bath modes with creation
and annihilation operators $b_k^\dagger$ and $b_k$ and coupling
strengths $g_k$.
The bath influence is encoded in the spectral~density
\begin{equation}
  J(\omega) = \sum_k |g_k|^2 \delta(\omega - \omega_k).
\end{equation}

When the spectral density can be approximated by a Lorentzian function,
as is common in microwave cavity systems \cite{Blais2021} and
nanophotonic environments \cite{Chikkaraddy2016},
\begin{equation}
  J(\omega) = \frac{1}{\pi} \frac{g^2 \kappa}{(\omega
    - \omega_a)^2 + \kappa^2},
\end{equation}
the pseudomode formalism shows that the exact reduced dynamics of the
system can be obtained by coupling it to a single damped harmonic
oscillator, or pseudomode, with frequency $\omega_a$, coupling strength
$g$, and damping rate $\kappa$ \cite{Garraway1997}.
The total Hamiltonian of the combined system and pseudomode is
\begin{equation}
  H_{pm} = H_S \otimes \mathbb{I} + \mathbb{I} \otimes \omega_a a^\dagger a
  + g \left(L \otimes a^\dagger + L^\dagger \otimes a \right),
\end{equation}
where $a$ and $a^\dagger$ are the annihilation and creation operators of
the pseudomode.
The evolution of the joint density operator $\rho(t)$ is governed by the
Markovian master equation
\begin{equation}
  \frac{d\rho}{dt} = -i[H_{pm}, \rho] + \kappa \left( a \rho a^\dagger
    - \frac{1}{2} \{ a^\dagger a, \rho \} \right).
\label{non}
\end{equation}

This representation effectively maps a non-Markovian open system problem
onto an extended Markovian one, enabling the use of Lindblad-form master
equations to simulate memory effects without explicitly dealing with
integrodifferential equations or memory kernels.
The parameter $\kappa$ determines the width of the Lorentzian spectral
density and thus controls the bath correlation time: small $\kappa$
corresponds to strong non-Markovianity with long bath memory (as
observed in high-Q cavities \cite{Leghtas2015}), while large $\kappa$
recovers the Markovian limit with rapid environmental decoherence.
Recent work has extended this approach to multiple pseudomodes for
complex spectral densities \cite{Guarnieri2018} and fermionic
environments \cite{Tamascelli2019}, demonstrating its versatility in
modeling modern quantum devices where environmental memory effects are
significant \cite{Cai2021}.

To explore the role of environmental memory in entanglement generation, we present a preliminary study of non-Markovian dynamics using the pseudomode formalism. By comparing protocols P1 and P2, we examine how memory effects influence the maximum entanglement achieved—effects that are typically neglected under the Markovian assumption.

We consider a single bosonic reservoir characterized by a Lorentzian spectral density, a widely used model that captures finite environmental correlation times. The system--reservoir coupling is implemented via the operator $L = \sum_{i=1}^{N} (S_i^z + S_i^x)$, which accounts for both dephasing and dissipative processes in a zero-temperature bath. The system Hamiltonian $H_S$ is given by Equation~\eqref{Ham}.

We note that richer non-Markovian features may emerge in the long-time dynamics. However, these effects lie beyond the scope of the present analysis, which is focused on fast, on-demand entanglement generation.

All numerical simulations in this work were performed using the QuTiP library~\cite{qutip2}. Equation~\eqref{non} was solved via QuTiP’s Monte Carlo solver \texttt{mcsolve}, with the number of trajectories chosen to ensure convergence. To balance the accuracy and computational cost, we adopted an adaptive Fock basis and validated its precision by evaluating the bosonic commutator $[a, a^\dagger]$, maintaining a numerical error below $10^{-4}$. Additionally, the convergence of all physical observables was confirmed through comparisons with larger basis sizes during preliminary tests.

{Figure}~\ref{fig:fig9} presents heatmaps of the maximum normalized end-to-end negativity achieved within an evolution time equal to twice the optimal entanglement generation time of the corresponding closed system, as a function of the system–reservoir coupling strength $g$ and the reservoir central frequency $\omega_a$. The three columns correspond to different values of the reservoir linewidth $\kappa$, which characterizes the spectral density of the environment and controls the degree of memory effects. The selected values—$\kappa = 0.01$, $\kappa = 1$, and $\kappa = 100$—represent, respectively, a strongly non-Markovian regime with long bath memory, an intermediate regime where the bath correlation times are comparable to the system dynamics, and the Markovian limit.

For $\kappa = 0.01$ and weak coupling ($g < 0.2$), P1 can only reach high entanglement when the bath frequency $\omega_a$ is significantly detuned from the system’s eigenfrequencies. In contrast, P2 maintains high entanglement over a broader range of $(g, \omega_a)$ values, demonstrating greater robustness to resonance-induced decoherence. However, for larger coupling strengths (e.g., $g \gtrsim 0.4$), both protocols fail to generate highly entangled states due to strong back-action from the environment.

When the bath linewidth is increased to $\kappa = 1$, P1 exhibits similar behavior to the previous case, remaining highly sensitive to $\omega_a$ and only achieving strong entanglement in narrow regions of the parameter space. P2, on the other hand, shows an expanded region of high negativity, further confirming its resilience to moderate non-Markovian noise.

In the Markovian limit ($\kappa = 100$), both protocols recover the results obtained using the Lindblad master equation in the previous section for $g \lesssim 0.4$, as expected. Notably, P2 continues to outperform P1 for stronger coupling values, retaining its advantage even as the system--reservoir interaction increases. These results highlight the superior robustness of P2 in the presence of both Markovian and non-Markovian environmental noise.

}
\begin{figure}[H]

  \includegraphics[width=\textwidth]{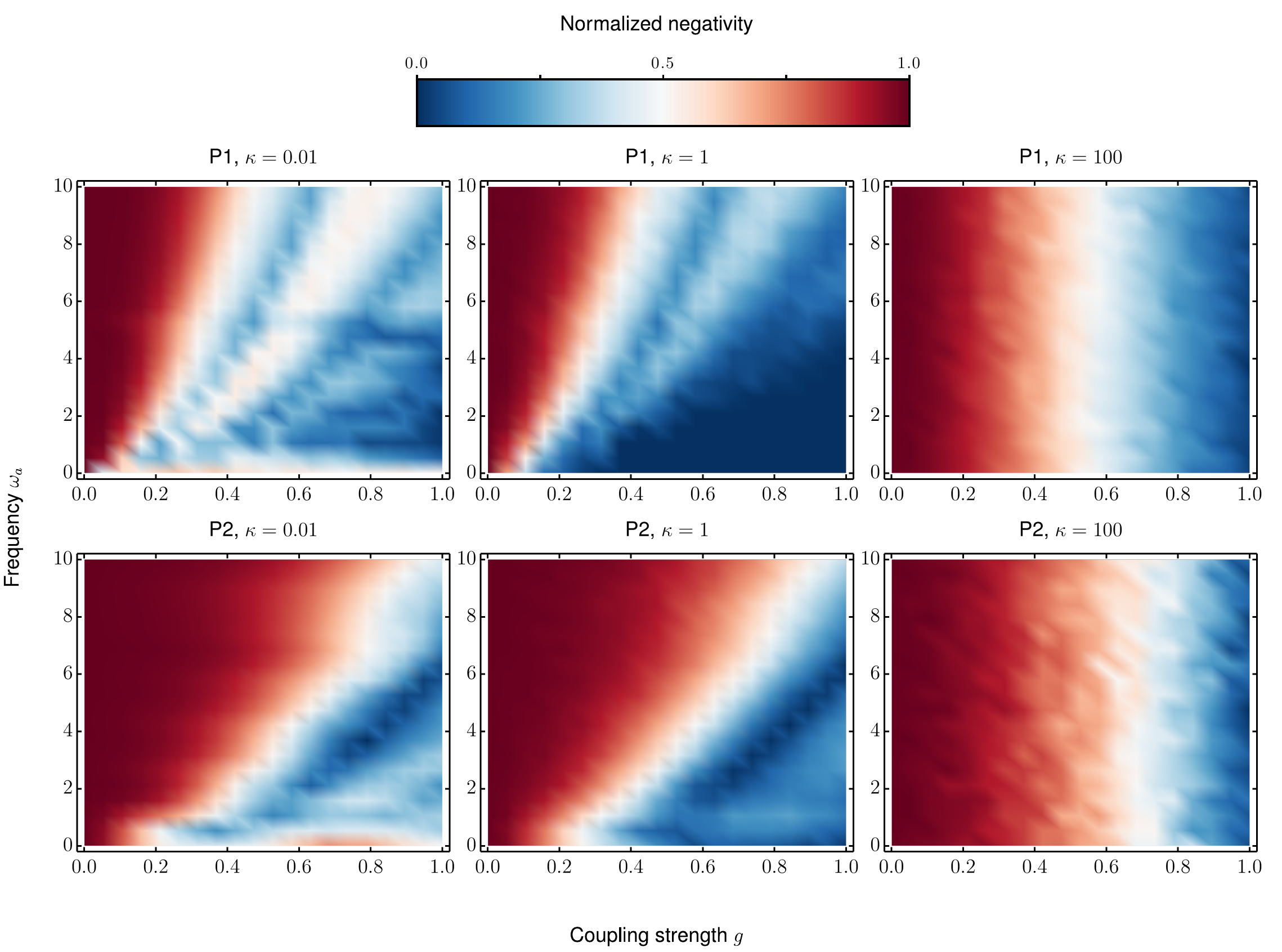}
  \caption{
    {Maximum} normalized end-to-end negativity achieved within an
    evolution time equal to twice the optimal transfer time of the
    closed system, for both protocols under non-Markovian dissipation
    modeled via the pseudomode method.
    Each column corresponds to a different spectral width $\kappa$ of
    the Lorentzian reservoir:
    (left) $\kappa = 0.01$ (strongly non-Markovian), (center) $\kappa =
    1$ (intermediate memory), and (right) $\kappa = 100$ (Markovian
    limit).
    The top row displays results for P1 and the bottom row for P2.
    For small $g$ and $\kappa$, P2 sustains high entanglement over a
    broader parameter region compared to P1.
    As $g$ increases, both protocols deteriorate, but P2 remains more
    resilient.
    In the Markovian limit, both recover the Lindblad results,
    validating the approach.
  }
  \label{fig:fig9}
\end{figure}
We emphasize that this is a preliminary analysis based on a simple yet physically relevant Lorentzian spectral density. More complex environments—such as sub-ohmic, super-ohmic, or multi-peaked spectra—would require additional pseudomodes to accurately capture their structures and may introduce qualitatively new dynamical features. A more detailed investigation of the full entanglement dynamics, beyond the short-time window associated with the maximum entanglement in the closed system, remains an important direction for future work.

\section{Conclusions}

We have conducted a comprehensive numerical study of two entanglement
generation protocols in XX spin chains, evaluating their performance
across spin magnitudes $s = {1}/{2}$, $1$, and ${3}/{2}$.
Our analysis reveals that the dual-port architecture (P2) consistently
achieves higher entanglement in shorter timescales than its staggered
counterpart (P1) for all spin values considered.

In addition to its speed advantage, P2 demonstrates strong robustness
against both diagonal and off-diagonal disorder, as well as local
dephasing noise.
This resilience is attributed to its design, which minimizes the excitation
of the bulk spins through optimized boundary control and coupling
symmetry, thereby enhancing the coherence and reducing the vulnerability to
noise.
\revision{
We also extended our analysis to incorporate non-Markovian
environmental effects using the pseudomode formalism.
This approach allowed us to assess how environmental memory, arising from
structured reservoirs with finite correlation times, affects the
entanglement dynamics of both protocols.
Our results show that P2 remains more robust than P1 across a wide range
of spectral parameters, further confirming its suitability for
implementation in noisy intermediate-scale quantum devices.
These findings open up promising directions for future studies of long-time
dynamics and more complex environments involving multiple pseudomodes
and structured spectral densities.}
These features make P2 not only more efficient but also more scalable
and robust, establishing it as a strong candidate for practical
implementations in entanglement-based quantum information processing.
Future work may explore the extension of this protocol to larger spin
networks, \revision{the incorporation of more general non-Markovian
effects \cite{de2017dynamics}}, and embedding it within hybrid quantum
architectures \cite{forn2019ultrastrong}.
These directions could be further enriched by integrating advanced
techniques such as color-engineered communication channels
\cite{de2013colored}, star-like entanglement hubs for multi-qubit
interfacing \cite{grimaudo2022spin}, and dissipative stabilization
mechanisms for steady-state entanglement \cite{de2017steady}.

\revision{
The XX spin chain with boundary magnetic fields studied in this work can
be robustly implemented using several state-of-the-art quantum
simulation platforms.
In circuit QED architectures \cite{Blais2021}, nearest-neighbor XX interactions
arise naturally between superconducting qubits (either flux or transmon
types) dispersively coupled to a shared microwave resonator, while
boundary $Z$-fields can be precisely engineered via local flux bias
lines or microwave drives.
Trapped-ion quantum simulators \cite{experiment2,experiment3} offer an
equally powerful alternative:
phonon-mediated interactions, induced by laser fields, generate
effective XX couplings, and site-resolved $Z$-fields can be realized
through differential AC Stark shifts or magnetic field gradients.

More generally, our model naturally emerges in systems of $N$ qubits
dispersively coupled to a common bosonic mode via Jaynes–Cummings-type
interactions \cite{zueco2009qubit}.
In such setups, the effective $Z$-field is set by the qubit detunings,
while spin-exchange interactions are mediated by virtual excitations of
the bosonic mode.
Crucially, this architecture enables full control over both the strength
and spatial profile of the spin–spin couplings, by tuning the individual
qubit mode detunings and coupling constants.
As a result, arbitrary patterns of exchange interactions—such as the
position-dependent profiles used in our protocols—can be engineered with
high precision.

These experimentally established platforms, each offering complementary
strengths in terms of coherence, control, and scalability, provide realistic and
versatile routes by which to test our predictions and implement the proposed
entanglement generation schemes in near-term quantum devices.
}

\vspace{6pt}
\authorcontributions{Conceptualization, E.K.S., G.D.d.M.N. and F.M.A.; Methodology, E.K.S., G.D.d.M.N. and F.M.A.; Formal analysis, E.K.S., G.D.d.M.N. and F.M.A.; Investigation, E.K.S., G.D.d.M.N. and F.M.A.; Writing---original draft, E.K.S. and G.D.d.M.N.; Writing---review \& editing, E.K.S., G.D.d.M.N. and F.M.A.; Supervision, F.M.A. All authors have read and agreed to the published version of the manuscript.
}

\funding{{This}
 work was partially supported by the Coordenação de Aperfeiçoamento de
Pessoal de Nível Superior (CAPES, Finance Code 001).
It was also supported by the Conselho Nacional de Desenvolvimento Científico
Tecnológico (CNPq) and Instituto Nacional de Ciência e Tecnologia de
Informação Quântica (INCT-IQ). F.M.A. acknowledges financial support from
CNPq Grant No. 313124/2023-0.}

\institutionalreview{{Not applicable.}
}

\dataavailability{{The datasets generated during this research are available from the corresponding authors upon reasonable request.}

}

\conflictsofinterest{The authors declare no conflicts of interest.
}

\appendixtitles{yes}
\appendixstart
\appendix
\section{Effective Dynamics}
\label{app:effective_dynamics}
To complement the numerical results discussed in the main text, we
present here a concise derivation of effective descriptions in two
relevant limiting regimes that support entanglement generation between
distant spins-${1}/{2}$: (i) the dispersive regime of P2 and
(ii) the strong dimerization limit, where the bulk dynamics are well
described by a trimer model involving only a few sites.

\subsection{Effective Dispersive Hamiltonian for Protocol 2}

We consider an $N$-site spin-$1/2$ XX chain with alternating couplings,
described by
\begin{equation}
  H = \sum_{i=1}^{N-1} J_i \left( S_i^x S_{i+1}^x
    + S_i^y S_{i+1}^y \right) + \sum_{i=1}^N B_i S_i^z,
\end{equation}
where $J_i = \Delta$ (strong) or $\delta$ (weak), and $B_i = \omega/2$
is a uniform magnetic field.
We consider emitter ($e$) and receiver ($r$) qubits coupled at the
boundaries via
\begin{equation}
  H_{\text{er}} = \frac{\omega}{2} (S_e^z + S_r^z)
  + \lambda \left( S_e^+ S_1^- + S_r^+ S_N^- + \text{H.c.} \right),
\end{equation}
leading to the full Hamiltonian $H _S= H + H_{\text{er}}$, with all
$J_i = \Delta$ and $\lambda = \delta$ for P2.

Applying the Jordan--Wigner transformation maps the spin system to
non-interacting fermions.
The resulting Hamiltonian in the single-excitation subspace becomes
\begin{equation}
  H'_S = \sum_{k=1}^N E_k f_k^\dagger f_k +
  \sum_{k=1}^N \overline{\lambda}_k \left( c_e^\dagger f_k
    + (-1)^{k-1} c_r^\dagger f_k + \text{H.c.} \right),
\end{equation}
where
\begin{align}
  E_k = {}
  & \Omega + 2 \Delta \cos\left( \frac{k \pi}{N+1} \right), \\
  \overline{\lambda}_k = {}
  & \lambda \sqrt{\frac{2}{N+1}} \sin\left( \frac{k \pi}{N+1} \right),
    \label{eq:lambdak}
\end{align}

{Transforming} to the interaction picture concerning
$H_0 = \sum_k E_k f_k^\dagger f_k + \frac{\omega}{2}(S_e^z + S_r^z)$
gives
\begin{equation}
  \bar{H}_S(t) =
  \sum_k \overline{\lambda}_k
  \left[
    \left( c_e^\dagger f_k
      + (-1)^{k-1} c_r^\dagger f_k \right) e^{i \zeta_k t}
    + \text{H.c.}
  \right],
\end{equation}
with detuning $\zeta_k = \omega - E_k$.

In the dispersive regime $\overline{\lambda}_k / \zeta_k \ll 1$,
second-order perturbation theory yields an effective Hamiltonian that
directly couples the boundary qubits via virtual excitations:
\begin{equation}
H_{\text{eff}} = \chi \left( c_e^\dagger c_r + c_e c_r^\dagger \right),
\end{equation}
with
\begin{equation}
\chi = \sum_k \frac{\overline{\lambda}_k^2}{\zeta_k}.
\end{equation}

This mediated interaction allows coherent entanglement generation
without significant excitation of the channel.
The average number of excitations in the chain during evolution up to
the entanglement generation time $\tau = \pi/4\chi$ is
\begin{equation}
  \left\langle n \right\rangle
  = \sum_k \int_0^\tau \frac{\left\langle f_k^\dagger(t) f_k(t)
    \right\rangle}{\tau} dt
  \approx N \left( \frac{\pi \delta}{2\Delta} \right)^2.
\end{equation}
Thus, the excitation leakage remains negligible as long as
$\sqrt{N}\delta / \Delta \ll 1$.

Moreover, the third-order corrections scale as
$N \overline{\lambda}_k^3 / \zeta_k^2$, so the validity of the effective
model requires
\begin{equation}
N \ll \frac{\zeta_k}{\overline{\lambda}_k}.
\end{equation}

\subsection{Trimer Approximation in the Strong Dimerization Regime}
In the opposite limit of static spin chains with alternating strong and
weak couplings, the $7$-site ABC spin chain can be reduced to an
effective three-site (trimer) model involving only the sites labeled
$A$, $B$, and $C$.
For $\delta / \Delta \ll 1$, the chain effectively breaks into dimers
weakly coupled via $\delta$. In this regime, the dynamics relevant to
entanglement generation can be captured by an effective trimer model
involving only the boundary spins $A$, $C$ and the central site $B$.

The effective Hamiltonian projected to the single-excitation subspace is
\begin{equation}
  H_{\mathrm{trimer}} =
  \begin{pmatrix}
    0    & \eta & 0 \\
    \eta & 0    & \eta \\
    0    & \eta & 0
  \end{pmatrix},
\end{equation}
with effective coupling
\begin{equation}
  \eta =
  \frac{\Delta}{2} \sqrt{1 + 3 \left( \frac{\delta}{\Delta} \right)^2
    - \sqrt{1 + 6 \left( \frac{\delta}{\Delta} \right)^2
      + \left( \frac{\delta}{\Delta} \right)^4 }}.
\end{equation}

This trimer mediates coherent oscillations between $A$ and $C$ through
$B$, enabling the generation of high-fidelity entanglement.
The first entanglement peak appears approximately at
\begin{equation}
  t_E \approx \frac{t_F}{2} = \frac{\pi}{2\sqrt{2} \eta},
\end{equation}
with the revival of the initial state at $t_F = \pi / \sqrt{2} \eta$.

This approximation generalizes to longer chains by symmetrically adding
dimer pairs around the central site $B$, preserving low-energy
trimer-like dynamics.
However, the effective coupling $\eta$ decreases exponentially with
the chain length, leading to a corresponding increase in the entanglement
timescale.

Both protocols allow for controlled entanglement generation between
distant qubits by exploiting virtual excitation pathways.
The effective Hamiltonians derived here enable fast, robust entanglement
in the presence of weak system--bath interactions and minimal occupation
of the intermediate chain.

\begin{adjustwidth}{-\extralength}{0cm}

\reftitle{{References}}

\PublishersNote{}
\end{adjustwidth}

\end{document}